\pdfoutput=1
\documentclass[conference]{IEEEtran}
\usepackage{cite}

\usepackage{amsmath,amssymb,amsfonts}
\usepackage{algorithmic}
\usepackage{graphicx}
\usepackage{textcomp}
\usepackage{xcolor}
\usepackage{csquotes}
\usepackage{flushend}
\usepackage{tikz}
\usetikzlibrary{svg.path}

\PassOptionsToPackage{hyphens}{url}
\usepackage{hyperref}
\def\BibTeX{{\rm B\kern-.05em{\sc i\kern-.025em b}\kern-.08em
    T\kern-.1667em\lower.7ex\hbox{E}\kern-.125emX}}

\newcommand\copyrighttext{%
  \footnotesize \textcopyright~2023 IEEE. Personal use of this material is permitted.
  Permission from IEEE must be obtained for all other uses, in any current or future
  media, including reprinting/republishing this material for advertising or promotional
  purposes, creating new collective works, for resale or redistribution to servers or
  lists, or reuse of any copyrighted component of this work in other works.
  }
\newcommand\copyrightnotice{%
\begin{tikzpicture}[remember picture,overlay]
\node[anchor=south,yshift=20pt] at (current page.south) {\fbox{\parbox{\dimexpr\textwidth-\fboxsep-\fboxrule\relax}{\copyrighttext}}};
\end{tikzpicture}%
}
    
\begin{document}

\title{Sources of Opacity in Computer Systems:\\Towards a Comprehensive Taxonomy}

\author{\IEEEauthorblockN{
Sara Mann\IEEEauthorrefmark{1},
Barnaby Crook\IEEEauthorrefmark{2},
Lena K\"{a}stner\IEEEauthorrefmark{2},
Astrid Schom\"{a}cker\IEEEauthorrefmark{2},
Timo Speith\IEEEauthorrefmark{2}\IEEEauthorrefmark{3}}
\IEEEauthorblockA{\IEEEauthorrefmark{1}Technical University Dortmund, Institute for Philosophy and Political Science, Dortmund, Germany}
\IEEEauthorblockA{\IEEEauthorrefmark{2}University of Bayreuth, Department of Philosophy, Bayreuth, Germany}
\IEEEauthorblockA{\IEEEauthorrefmark{3}Saarland University, Center for Perspicuous Computing, Saarbr\"{u}cken, Germany}

Email: sara.mann@tu-dortmund.de, \{barnaby.crook, lena.kaestner, astrid.schomaecker, timo.speith\}@uni-bayreuth.de
}

\maketitle

\copyrightnotice
\vspace{-2ex}

\begin{abstract}
Modern computer systems are ubiquitous in contemporary life yet many of them remain opaque. This poses significant challenges in domains where desiderata such as fairness or accountability are crucial. We suggest that the best strategy for achieving system transparency varies depending on the specific \emph{source} of opacity prevalent in a given context. Synthesizing and extending existing discussions, we propose a taxonomy consisting of eight sources of opacity that fall into three main categories: architectural, analytical, and socio-technical. For each source, we provide initial suggestions as to how to address the resulting opacity in practice. The taxonomy provides a starting point for requirements engineers and other practitioners to understand contextually prevalent sources of opacity, and to select or develop appropriate strategies for overcoming them.
\end{abstract}


\section{Introduction}
\label{sec:intro}

Computer systems are omnipresent in modern life, ranging from the smartphones in our pockets, to computer simulations in science, to artificial intelligence (AI) systems utilized in the private sector. Many of the systems in question are \emph{opaque}, i.e., there exists some barrier hindering our understanding of how exactly they work. This is problematic, as it is widely acknowledged that opacity can hinder the fulfillment of a broad range of societal desiderata (e.g., scientific progress or fairness) \cite{Langer2021What, BarredoArrieta2020Explainable, Lipton2018Mythos, Chazette2021Exploring, Koehl2019Explainability, Speith2022Review}, especially in high-stakes scenarios such as medical diagnosis \cite{Lundervold2019Overview} or autonomous driving \cite{Grigorescu2020Survey}.\footnote{It should be noted that opacity is not always problematic \cite{Humphreys2009Philosophical, Beisbart2021Opacity} and can even be beneficial, at least in some situations and to some stakeholders \cite{Langer2023Introducing}.} 

Against this backdrop, researchers are developing strategies that seek to remove the barriers to understanding that make a given system opaque, e.g., by making it explainable to various stakeholders \cite{Langer2021What, Chazette2021Exploring, Koehl2019Explainability}. However, a variety of contextual factors influences whether an approach to addressing opacity succeeds \cite{Langer2021What}. In light of this, we suggest that the \emph{source} of opacity crucially affects what the best means to tackle a system's opacity will be in a given context. 

Although there is a vast body of literature on the opacity of different computer systems, such as AI systems \cite{Burrell2016Machine, Facchini2022Towards} and scientific computer simulations \cite{Humphreys2009Philosophical, Beisbart2021Opacity, Kaminski2018Mathematische}, we observe two problems with existing accounts: First, extant literature identifies several sources of opacity, but a \emph{comprehensive} taxonomy of potential sources of opacity in different kinds of computer systems is still unavailable. Second, most authors try to provide a theoretical description of opacity, while \emph{actionable suggestions} for how to address its sources are scarce.

We set out to alleviate these problems. Synthesizing and augmenting existing discussions, we distinguish eight sources of opacity, which we categorize as either \emph{architectural}, \emph{analytical}, or \emph{socio-technical}. For each of these sources, we propose initial strategies for how they could be addressed by requirements engineers and other practitioners.

We begin this vision paper by reviewing existing accounts of opacity in computer systems (\autoref{sec:opacity}). Integrating and expanding these accounts, we outline our taxonomy and suggest measures that can be taken to reduce opacity depending on its source (\autoref{sec:sources}). We conclude the paper by outlining some avenues for future research
(\autoref{sec:outlook}).

\section{Opacity in Computer Systems}
\label{sec:opacity}

We take (an aspect of) a computer system to be opaque if there exists some barrier to achieving knowledge or understanding about it (see Beisbart's account of opacity \cite{Beisbart2021Opacity};  \cite{Humphreys2009Philosophical, Alvarado2021Explaining} provide further definitions). This barrier can have an objective dimension (e.g., a deep neural network is objectively more opaque than a hand-coded rule-based system), but can also be stakeholder-dependent (e.g., a system can be transparent for an expert but opaque for a layperson) \cite{Alvarado2021Explaining}.

Opacity has gained attention in the literature on different kinds of computer systems \cite{Burrell2016Machine, Facchini2022Towards, Humphreys2009Philosophical, Beisbart2021Opacity, Kaminski2018Mathematische, Alvarado2021Explaining, Jebeile2018Collaborative, Alvarado2017Big}. While previous work provides valuable insights on opacity, we argue that extant suggestions lack completeness and/or actionability.

Beginning with a lack of \emph{completeness}, a comprehensive overview uniting all potential sources of opacity is still unavailable. Some authors focus only on one specific aspect of opacity. Humphreys, for instance, emphasizes its subjective dimension by defining opacity either relative to the agent's knowledge or their nature \cite{Humphreys2009Philosophical}. This definition does not capture that overcoming an agent's limitations is not always the best way to address opacity, especially in objectively highly opaque systems \cite{Alvarado2021Explaining}. Jebeile, on the other hand, discusses opacity due to the division of labor in scientific collaborations, and how this opacity is exacerbated by the use of technologies and by institutional secrecy with respect to data and the functioning of scientific instruments \cite{Jebeile2018Collaborative}. Still, she is not concerned with opacity that exists independently of social contexts \cite{Jebeile2018Collaborative}. 

Instead of emphasizing one aspect of opacity, several papers stress that opacity can stem from a variety of sources \cite{Burrell2016Machine, Facchini2022Towards, Alvarado2017Big, Kaminski2018Mathematische}. However, each of them discusses a slightly different selection of sources. For instance, among these papers, only the work of Kaminski et al. \cite{Kaminski2018Mathematische} takes the above-mentioned opacity due to division of labor into account. At the same time, they do not discuss other sources of opacity, such as \emph{intentional concealment} due to private or public actors keeping the functioning of computer systems secret. This source, however, is acknowledged by other authors (e.g., \cite{Burrell2016Machine, Facchini2022Towards}). 

Furthermore, we integrate three additional sources of opacity that are not recognized by any of these overviews (viz., \emph{lost knowledge}, v. \cite{Arbesman2017Overcomplicated}; \emph{missing tools}; and \emph{lacking resources}, v. \cite{Beisbart2021Opacity}). As a consequence, none of the existing works comprises all of the above sources of opacity in computer systems.

Moving to \emph{actionability}, suggestions on how to overcome opacity depending on its source are rarely made. Notable exceptions are Langer and K{\"o}nig \cite{Langer2023Introducing} and Burrell \cite{Burrell2016Machine}. These authors propose several strategies to deal with opacity depending on its source (e.g., regulation, education, or explainability). However, as both papers distinguish only three sources of opacity, the solutions they suggest address only a subset of potential sources of opacity in computer systems.

\section{Sources of Opacity}
\label{sec:sources}

\begin{figure*}
    \centering
    \includegraphics[width=\textwidth]{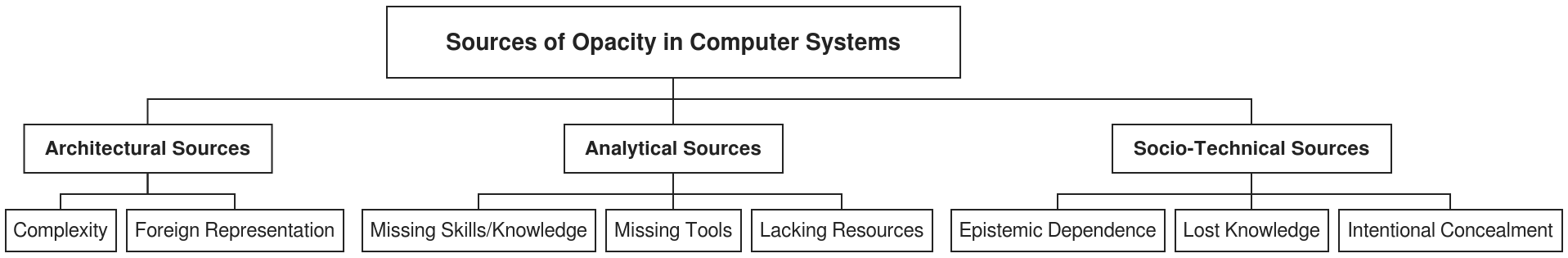}
    \caption{Our taxonomy of sources of opacity in computer systems.}
    \label{fig:taxonomy}
    \vspace{-2.2ex}
\end{figure*}

In this section, we aim to alleviate the problems identified above by providing a more comprehensive taxonomy of eight sources of opacity, each complemented by initial proposals for actionable strategies to address it. To this end, we divide the sources into three main categories: \emph{architectural}, \emph{analytical} and \emph{socio-technical} opacity (see \autoref{fig:taxonomy}). Depending on which aspect is emphasized, some sources might be classified in different ways. We mention these ambiguities in the respective sections. Importantly, the same system can be affected by multiple sources simultaneously. For concision, we choose not to distinguish different system-related aspects or levels of abstraction (for discussions see \cite{Facchini2022Towards, Creel2020Transparency, Langer2023Introducing, Beisbart2021Opacity}).

\subsection{Architectural Opacity}

\emph{Architectural} opacity is opacity that stems from the structural properties of computer systems. It may arise from the complexity of these systems or may be traced to how these systems represent and process information. 

\subsubsection{Complexity}

The complexity of computer systems is a source of opacity that is frequently emphasized, but rarely specified. In theoretical computer science, computational complexity refers to the computational resources required to solve a problem. We will return to this aspect when discussing \emph{limited resources} (see \autoref{sec:limres}).

Computer systems, and especially AI systems, are also complex in the colloquial sense of the word. In this more general sense, the complexity of any system can be characterized by two dimensions: i) the number of the system's elements\footnote{What is considered a distinct element of a complex system is a contextual question. For instance, understanding neural networks on the level of neurons might be useful for some computer scientists while unsuitable for end users.}, and ii) their interactions \cite{Arbesman2017Overcomplicated}.

On the one hand, a system can be complex because of scale. For example, a program consisting of hundreds of lines of code cannot be contemplated all at once \cite{Alvarado2021Explaining}. Similarly, a decision tree (or one of its decisions) can be complex due to a large number of nodes \cite{Lipton2018Mythos, VanOtterlo2013Machine}. This issue is especially severe for the millions of parameters of some neural networks. A variant of this opacity emerges when the relevant information itself is not complex, but is obscured by a vast quantity of surrounding information \cite{Ananny2018Seeing}. If this is done on purpose \cite{Ananny2018Seeing}, it is better described as \emph{intentional concealment} (see \autoref{sec:secrecy}). 

On the other hand, the interaction of the system's elements can lead to complexity. Non-linearity is one of the most common properties of complex systems \cite{Davidsson2017Simulation} and is also present in some AI models (e.g., in the activation functions of neural networks). Another type of interaction that gives rise to complexity is feedback loops \cite{Davidsson2017Simulation} which are also present in the context of AI (e.g., in recurrent neural networks).\footnote{Analysis of further properties of complex systems (e.g., emergence, self-organization) is beyond the scope of the paper. For discussion, see \cite{Strevens2003Bigger, Ladyman2013What, Mitchell2009Complexity, Davidsson2017Simulation}.}

A complex interaction of system elements can also occur in traditional software. For instance, so-called \emph{kludges} can lead to the (partial) opacity of computer code \cite{Creel2020Transparency, Lenhard2010Holism}. Kludges are unconventional makeshift solutions or workarounds to a programming problem. While effective in the short run, kludges can lead to problems down the line as they are often poorly understood and behave in unexpected ways \cite{Clark1987Kludge}. As kludges often arise during collaborative and iterative software development \cite{Lenhard2010Holism} they can lead to \emph{epistemic dependence} (see \autoref{sec:doel}) and are susceptible to \emph{lost knowledge} (see \autoref{sec:lostknow}). More generally, opacity due to complexity can also be seen as stemming from individual or human cognitive limits, rendering it an analytical issue (see \autoref{sec:anis}). 

\emph{Strategies:} The complexity of a model or software can be reduced with methods like feature extraction or by following best practice standards of programming \cite{Burrell2016Machine}. Explainability approaches that make use of deliberate simplification can also render a complex system tractable \cite{Ribeiro2016Why}. Any explanation that selects contextually relevant pieces of information can be seen as providing such a simplification (e.g., a saliency map highlighting image regions decisive for a classification \cite{Simonyan2013Deep}, or a counterfactual statement describing a way of altering the input to receive a different output \cite{Wachter2017Counterfactual}).

\subsubsection{Foreign Representation}

Sometimes, opacity has to do with the representations a computer system relies on. In particular, the way a system represents information or the representation's content can be foreign to humans. Both are mainly of concern in AI systems that rely on distributed, sub-symbolic representations that were autonomously generated during training, instead of being coded by hand \cite{Goodfellow2016Deep}.

On the one hand, a distributed representation might correspond to a symbolic representation familiar to humans (e.g., if a neural network learns a pattern that matches the human concept of \enquote{tree} \cite{Bau2017Network}). In this case, only the \emph{way} of representing information is foreign. On the other hand, research on adversarial examples suggests that some AI systems respond to subtle patterns that are meaningless to humans \cite{Szegedy2014Intriguing}. In such cases, not only the way of representing information, but also the representation's \emph{content} is foreign.
Notice that in both cases, a representation is foreign \emph{to humans}. As such, this source might be considered to also have an analytical component (see \autoref{sec:anis}).

\emph{Strategies:} In the first case, opacity can be overcome by mapping the distributed representations to human concepts. This can be done with explainable AI (XAI) techniques like saliency maps \cite{Simonyan2013Deep}, network dissection \cite{Bau2017Network}, and concept activation vectors \cite{Kim2018Interpretability}. In the second case, transparency might be achieved by introducing novel concepts into our language that capture the foreign representational content \cite{Schubert2021High-Low}. If it turns out that some of these representations are untranslatable to an intelligible concept, approximation might be a solution \cite{Baum2022Responsibility}.

Furthermore, so-called ante-hoc explainable AI systems \cite{Speith2022Review} are transparent in the sense that they avoid the way or content of representations being foreign from the start. Especially in high-stakes scenarios, they might be the best strategy to avoid opacity arising from foreign representations \cite{Burrell2016Machine, Rudin2019Stop}.

\subsubsection{Link Opacity} A related issue is so-called \enquote{link opacity} \cite{Facchini2022Towards} or \enquote{link uncertainty} \cite{Sullivan2022Understanding}. It occurs when it is unclear how the system can be used to explain phenomena---e.g., because it is uncertain which parts of the model represent which parts of the real world  \cite{Facchini2022Towards}. As this is a consequence of the combination of \emph{complexity} and \emph{foreign representation}, we do not list link opacity as a distinct source in \autoref{fig:taxonomy}. 

\emph{Strategies:} Link opacity can be approached either by addressing the system's architectural opacity as described above, or by gaining more knowledge about the target phenomenon \cite{Imbert2017Computer, Sullivan2022Understanding}. The latter can also be seen as acquiring \emph{missing skills and knowledge} (see \autoref{sec:missknow}).

\subsection{Analytical Opacity}
\label{sec:anis}

\emph{Analytical} opacity is opacity that stems from the absence of the skills, knowledge, and/or tools which would be necessary to analyze a computer system so as to render it transparent.

\subsubsection{Missing Skills and Background Knowledge}
\label{sec:missknow}

As noted by several authors, it is clear that the degree of opacity is mediated by individual knowledge and skill (e.g., programming) \cite{Facchini2022Towards, Burrell2016Machine, Alvarado2017Big, Jebeile2018Collaborative, Langer2023Introducing}. This is amplified by the rising application of sophisticated technology in a variety of domains involving several, diverse groups of stakeholders \cite{Langer2023Introducing, Langer2021What}. Within XAI, for instance, methods that are directed towards computer scientists prevail, neglecting the needs of domain experts or laypeople that use or are affected by AI systems \cite{Miller2017Explainable}. 

\emph{Strategies:} Opacity that stems from a lack of individual knowledge or skill can be addressed in several ways: by developing explainability approaches that are tailored to different stakeholder groups \cite{Langer2023Introducing, Langer2021What}, by obtaining skills and knowledge through education \cite{Langer2023Introducing, Burrell2016Machine}, or by accepting the opacity and deferring to experts instead (see \autoref{sec:doel}).

\subsubsection{Missing Tools}
\label{sec:misstool}

While possessing the relevant (e.g., computer science or domain) expertise is sufficient for understanding some computer systems, others can only be rendered explainable through the development and application of specific tools. If the tools required for addressing opacity in any given case have not been created or are unavailable, then the computer system will remain opaque. 

\emph{Strategies:} Since there are many kinds of computer systems, there are many tools which may be required to render them transparent. For example, XAI methods are tools to produce explanatory information about the decision-making procedures of AI systems \cite{Speith2022Review}. More generally, visualization techniques can be applied to AI systems (e.g., saliency maps) \cite{Molnar2019Interpretable}, traditional software systems (e.g., architecture graphs) \cite{Kienle2007Requirements}, and computer simulations (e.g., state trajectories) \cite{Jebeile2018Explaining} to increase their transparency. Further tools include searchable databases, programming languages \cite{Turbak2008Design}, and knowledge-sharing platforms (e.g., GitHub). Overall, the development and use of appropriate tools is crucial for addressing opacity.

\subsubsection{Lacking Resources}
\label{sec:limres}

In some cases, opacity may arise or persist because the resources required for transparency, such as computational power, time, or money, are not available \cite{Crook2023Revisiting, Beisbart2021Opacity}.
A lack of resources is closely linked to computational complexity as mentioned above: Computational complexity theory classifies computational problems into different complexity classes according to the computational resources required to solve them. The computational complexity of software in general and AI in particular \cite{Russell2021Artificial}, as well as of XAI approaches \cite{Molnar2019Interpretable}, is an important metric in computer science. 

\emph{Strategies:} This source of opacity can be addressed by either increasing the available resources to a sufficient degree (which might be impossible for computationally intractable problems) or by decreasing the resources required (e.g., by using a less computationally complex model or explainability approach).

\subsection{Socio-Technical Opacity}

\emph{Socio-technical} opacity is opacity that results from the social processes or arrangements that software systems are deployed in. This form of opacity cannot be attributed solely to the properties of software systems themselves, but also incorporates the human contexts in which they are embedded. 

\subsubsection{Epistemic Dependence}
\label{sec:doel}

Both creating from scratch and understanding complex computer systems often exceeds the skills and cognitive capabilities of an individual and requires the collaboration of multiple experts with different areas of expertise. Individuals are thus \emph{epistemically dependent} \cite{Hardwig1985Epistemic} upon their collaborators.

Epistemic dependence can be a source of opacity if an individual cannot assess their colleagues' contributions because they lack the required expertise \cite{Wagenknecht2014Opaque, Kaminski2018Mathematische, Jebeile2018Collaborative}. This is a common phenomenon in computer science, e.g., when programmers rely on programming languages developed by others instead of dealing with machine language \cite{Creel2020Transparency}, or on external libraries they treat as black boxes \cite{Mittelstadt2016Ethics}. These examples illustrate that epistemic dependence may be closely linked to specialization and thus to \emph{missing skills and knowledge} (see \autoref{sec:missknow}) and \emph{missing tools} (see \autoref{sec:misstool}).

\emph{Strategies:} Epistemic dependence can be resolved in principle (e.g., by acquiring all kinds of expertise required for achieving complete transparency) but persists in practice for pragmatic reasons, such as time constraints. At the same time, different experts can collaborate to address opacity collectively and to increase overall transparency, albeit perhaps not for the individual. For that reason, this source of opacity may have to be accepted as a necessary evil. Furthermore, deference to experts \cite{Sloman2016Your, Fricker2006Testimony} can be a strategy to deal with opacity for laypeople whose epistemic dependence is likely to persist. This requires trustworthy institutions, e.g., legislative bodies, NGOs or research institutes. 

\subsubsection{Lost Knowledge}
\label{sec:lostknow}

Many computer programs or programming languages are employed over extended periods of time. If nobody understands old programs and languages anymore, these become opaque due to \emph{lost knowledge} (see \cite{Arbesman2017Overcomplicated}). A prominent example is COBOL, a programming language that originated in the 1950s but is still widespread in business applications \cite{Ciborowska2021Contemporary}. As the last generation of COBOL programmers retires and COBOL has been out of most curricula for decades, companies now lack experts capable of maintaining legacy code \cite{Lindoo2014Bringing}. As lost knowledge is a form of lacking knowledge, it can also be considered an analytical issue (see \autoref{sec:anis}).

\emph{Strategies:} This source of opacity can be addressed by recovering lost knowledge (e.g., by reverse engineering) or replacing legacy systems with contemporary software. Furthermore, producing adequate documentation \cite{Parnas2010Precise} and educating trainees can preempt this source from occurring by preserving the knowledge before it is lost.

\subsubsection{Intentional Concealment}
\label{sec:secrecy}

In some cases, public or private institutions that develop or employ computer systems purposefully seek opacity. Opacity can be a means to exercise control \cite{Langer2023Introducing}, to escape regulation, or to hide problems and unlawful behavior \cite{Burrell2016Machine}. Intentional concealment can also happen for benevolent reasons (e.g., when user-friendly interface design  hides the underlying complexity of a computer system, or when governments keep their programs secret for security reasons). Other motivations for seeking opacity include preventing a system from being gamed \cite{Burrell2016Machine, Langer2023Introducing}, protecting intellectual property \cite{Alvarado2017Big, Langer2023Introducing}, as well as securing economic \cite{Burrell2016Machine, Facchini2022Towards, Langer2023Introducing} or scientific \cite{Jebeile2018Collaborative} advantage.

\emph{Strategies:} Intentional concealment can be addressed in several ways: by reverse engineering \cite{Burrell2016Machine}, as with the German credit scoring algorithm SCHUFA or the recidivism risk-scoring system COMPAS \cite{Rudin2020Age};\footnote{Regarding the former, see \url{https://blog.okfn.org/2018/11/29/openschufa-the-first-results/}; as for the latter, see \url{https://www.propublica.org/article/machine-bias-risk-assessments-in-criminal-sentencing}. Accessed (both): 11 July 2023.}  by law \cite{Burrell2016Machine}, as exemplified by the GDPR \cite{Goodman2017European, Langer2023Introducing}; by relying on open-source software instead \cite{Burrell2016Machine}; or even by whistle-blowing.

\section{Conclusions and Outlook}
\label{sec:outlook}

In this vision paper, we synthesized existing accounts of opacity into a taxonomy that goes beyond any individual existing approach. We identified eight sources of opacity in computer systems which can be broadly classed into architectural, analytical, and socio-technical opacity. For each source, we offered initial strategies practitioners might pursue to address the resulting opacity. However, our taxonomy is only a starting point to identify strategies for effectively addressing opacity. Our suggestions will need to be complemented, specified, and validated with practitioners including requirements engineers.

In particular, we think that employing the elements of our taxonomy in a workflow might be a fruitful way to identify present sources of opacity and select suitable means for addressing them. For instance, one could identify the source of opacity by asking first whether the computer system is inaccessible due to intentional concealment. In a second step, one could determine whether the computer system is opaque to relevant experts. If not, the opacity likely stems from the lack of skills, knowledge, or tools of laypersons. Otherwise, only sources of opacity that apply to experts remain (e.g., complexity, foreign representation, or epistemic dependence).

To implement such a workflow, further refinements of our proposal might be required. First, the various ways in which different sources of opacity interact and overlap, either theoretically or practically, need to be investigated. Second, it would be useful to establish taxonomies or workflows that are tailored towards the needs and prior knowledge of specific stakeholders \cite{Speith2022Review}. Third, more work is needed to identify and develop specific strategies for addressing each of the different sources of opacity in different contexts. Finally, discussion is needed to illuminate under which conditions and to which extent opacity may be beneficial (or at least acceptable) to not overburden individuals with information and thereby hinder their understanding, trust, and decision-making. We aim to address these issues in future work.

\section*{Acknowledgments}
Work on this paper was funded by the Volkswagen Foundation grants AZ 9B830, AZ 98510, and AZ 98514 \href{https://explainable-intelligent.systems}{\enquote{Explainable Intelligent Systems}} (EIS) and by the DFG grant 389792660 as part of \href{https://perspicuous-computing.science}{TRR~248}. We thank Eva Schmidt, Julian Speith, and three anonymous reviewers for their helpful feedback.

\bibliographystyle{IEEEtran}
\bibliography{bibliography}

\end{document}